\documentclass[showpacs,floatfix,twocolumn,prl]{revtex4}

\usepackage{graphicx,color}

\begin{document}

\title{Real space investigation of structural changes at the metal-insulator 
transition in VO$_2$}

\author{Serena A. Corr} \email{s.a.corr@kent.ac.uk}
\affiliation{School of Physical Sciences,
University of Kent, Canterbury, Kent CT2 7NH, UK}

\author{Daniel P. Shoemaker}
\author{Brent C. Melot}
\author{Ram Seshadri}
\affiliation{Materials Department and Materials Research Laboratory,
University of California, Santa Barbara, CA, 93106, USA}


\begin{abstract}
Synchrotron X-ray total scattering studies of structural changes in
rutile VO$_2$ at the metal-insulator transition 
temperature of 340\,K reveal that monoclinic and tetragonal phases 
of VO$_2$ coexist in equilibrium, as expected for a first-order phase
transition. No evidence for any distinct 
intermediate phase is seen. Unbiased local structure studies of the changes 
in V--V distances through the phase transition, using reverse Monte Carlo 
methods, support the idea of phase coexistence and point to the high degree 
of correlation in the dimerized low-temperature structure. No evidence for 
short range V--V correlations that would be suggestive of \textit{local} 
dimers is found in the metallic phase.
\end{abstract}


\maketitle 

\begin{figure}
\centering\includegraphics[width=0.7\columnwidth]{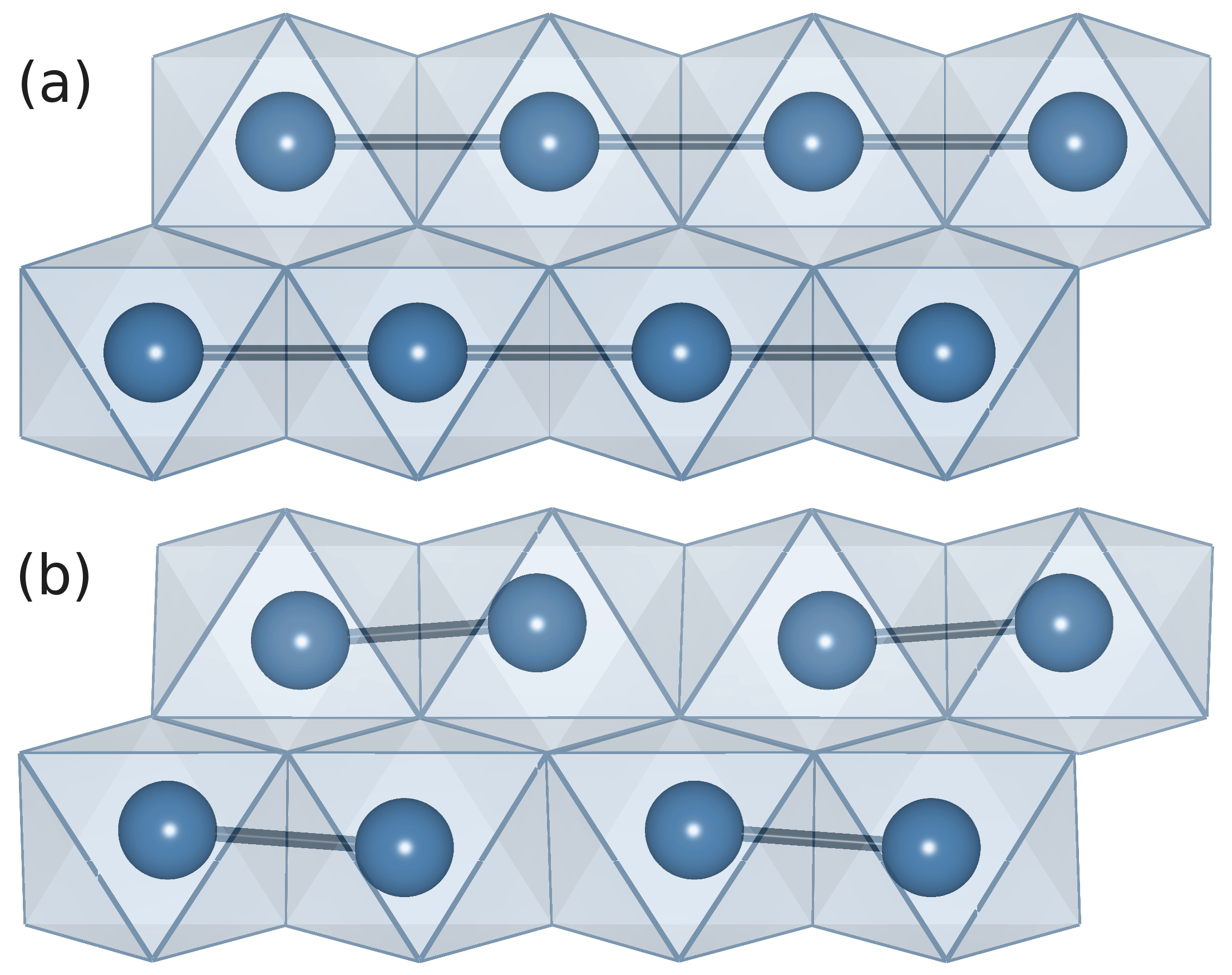} \\
\caption{(Color online) (a) The metallic high temperature tetragonal rutile
form of VO$_2$ with a single V--V distance and (b) the insulating
low-temperature  monoclinic form, showing dimerized chains of alternating
short and long V--V distances along the $c$-axis.}
\label{fig:chains}
\end{figure}

VO$_2$ undergoes a transition from a metal to a non-magnetic insulator upon 
cooling below 340\,K.\cite{Morin}  Accompanying this transition is a 
structural change from the high temperature tetragonal phase to a low 
temperature monoclinic phase, where pairing and tilting of vanadium ions 
result in chains with alternating long and short V--V distances along the 
$c$-axis (Fig.~\ref{fig:chains}).\cite{Goodenough1960,Berglund,Zylbersztejn}  
The mechanism underpinning the insulating behaviour of the low temperature 
phase continues to attract widespread interest, and recent work suggests 
the simultaneous role of the V--V pairing and electron correlation to drive
the system from a correlated metal to a slightly less correlated 
insulator.\cite{Imada,Biermann,Cava} 
Diverse methods have been used to understand this transition, including 
recent probes of dynamics and techniques sensitive to spatial 
inhomogeneities.\cite{Cavalleri,Qazilbash_Science,photoemission,zewail}
The phenomenon as it occurs in nanostructures is also receiving increasing
attention.\cite{Corr,Welland,Lauhon}

Above 340\,K, the rutile VO$_2$ structure in space group 
$P4_2/mnm$ has a single near-neighbor V--V distance of
2.88\,\AA.\cite{VO2_struc,xanes}  Below 340\,K, the monoclinic structure in 
space group $P2_1/c$ is characterized by 
V--V dimers along the $c$-axis displaying two distinct V--V distances: 
2.65\,\AA\/ and 3.12\,\AA\/.\cite{Andersson,xanes,SC2028}  The structures are 
displayed in Fig.\,\ref{fig:chains}.
Interestingly, several recent papers address the formation of domain-like 
regions in monoclinic VO$_2$ on approaching the transition 
temperature from below.\cite{domains1,Kim_phonon,Qazilbash_Science,Cava} 
An unresolved question regarding the nature of these domains 
is whether they are associated with an intermediate phase formed during
the transition. A second question concerns the possible existence of dimers, 
albeit over short length scales, persisting in the metallic high temperature phase.  

Here we employ synchrotron X-ray total scattering in conjunction with 
real-space analysis---using least-squares as well as reverse Monte Carlo 
methods---to show that the transition from the low-temperature monoclinic
to the  high temperature tetragonal phase in rutile VO$_2$ occurs in a
first-order manner, with coexistence of the low and high-temperature phases 
at the transition temperature. We employ pair distribution function (PDF) 
analysis, which has emerged in recent years as an indispensable tool for 
probing and understanding structural changes, and specifically, the 
range-dependence thereof.\cite{Billinge_LaMnO3,Petkov_PDF,Chupas_heat2,Kate_1} 
The advantages of the PDF method lie in its ability to probe structural 
changes that lack long range correlations, which are not captured by 
Bragg peaks in X-ray or neutron scattering. 
By analyzing PDF data with large-box reverse Monte Carlo (RMC) simulations,
we are able to  produce a supercell where the local environment is free from 
symmetry constraints of the high or low temperature 
structures.\cite{goodwin_ferroelectric_2007,shoemaker_unraveling_2009,shoemaker_atomic_2010}
Atomic 
positions can therefore relax to best fit the experimental data, which 
includes both the Bragg and diffuse components. This approach provides an 
opportunity to observe any continuum of atomic positions which might exist 
across the metal-to-insulator transition in VO$_2$.


Phase pure VO$_2$ powders were prepared in evacuated and sealed silica ampoules 
from a stoichiometric mixture of V$_2$O$_5$ (99.9\%, Alfa Aesar) and 
V$_2$O$_3$, the latter obtained by reduction of V$_2$O$_5$ in 5\%-H$_2$/N$_2$ gas at 
900$^{\circ}$C. Synchrotron total scattering data on powders in Kapton
tubes were collected in transmission mode at beamline 11-ID-B of the Advanced 
Photon Source at Argonne National Laboratory by using X-rays 
with energies near 90\,keV (wavelength $\lambda$ = 0.13702\,\AA). Samples were 
heated and cooled continuously from 250\,K to 400\,K at a rate of 
50\,K\,h$^{-1}$ using an Oxford Cryosystems Cryostream 700. Scattering data 
were collected on an image plate system with a sample-to-detector distance of 
250\,mm. Raw images were processed using \textsc{fit2d}.\cite{FIT2D} 
PDFs were extracted as $G(r)$ using
\textsc{pdfgetX2} with $Q_{max}$ = 25\,\AA$^{-1}$.\cite{pdfgetX2}  
Least-squares profile refinements were carried out 
using \textsc{pdfgui}.\cite{pdfgui} RMC simulations were performed using 
\textsc{rmcprofile}\cite{tucker_rmcprofile_2007} 
with a $10\times10\times16$ supercell, 
starting from atomic positions of the tetragonal structure. 
The simulations were constrained by the PDF (where $G(r)$ above corresponds
to $D(r)$ as 
described by Keen \cite{keen_comparison_2001}) up to $r$ = 25\,\AA. RMC
results shown in this work are averages of many simulations to 
ensure an unbiased representation of the fit to data.


\begin{figure}
\centering\includegraphics[width=0.9\columnwidth]{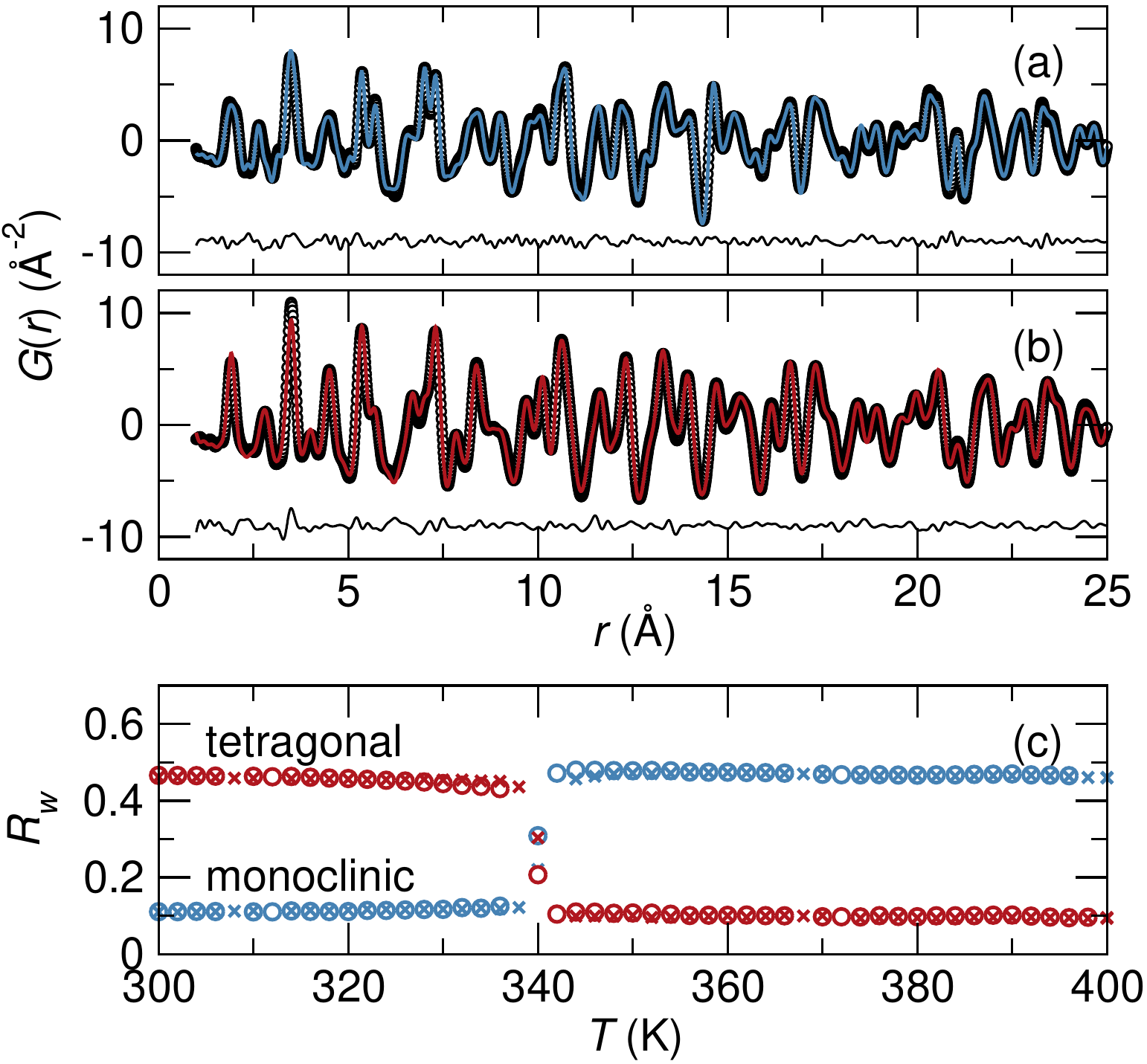} \\
\caption{(Color online) Least-squares fits to the PDF of bulk VO$_2$
using (a) the low temperature monoclinic structure for  
$T$ = 300\,K (R$_w$, 11.6\%) and (b) the high temperature tetragonal 
structure for $T$ = 400\,K  (R$_w$, 9.6\%). Circles are 
experimental data and lines are fits to the average structure. The difference 
is displayed below each fit. (c) Changes in the $R_w$ 
fitting parameters, shown for cooling ($\circ$) and heating ($\times$) through the transition 
temperature, show an abrupt transition with good fits at all temperatures
outside of $T$ = 340 K.}
\label{fig:pdfgui}
\end{figure}

Least-squares refinements to all the PDFs at 2\,K intervals 
were performed using both monoclinic and tetragonal structures 
[Fig.~\ref{fig:pdfgui}(a) and (b), 
respectively].  In the  case of data fit to a monoclinic structure 
[Fig.~\ref{fig:pdfgui}(a)], we obtain a good fit from $T$ =  250\,K to 
338\,K.  The high temperature tetragonal structure [Fig.~\ref{fig:pdfgui}(b)], 
results in a good fit to the data from $T$ = 342\,K to 400\,K. The PDFs of the 
two structures are markedly different as seen in Fig.\,\ref{fig:pdfgui}(a,b).
There are no significant changes (other than lattice constants) in either 
end member until the transition is reached. 
The sudden structural transition at $T$ = 340\,K is evident in plots of 
the goodness-of-fit parameters $R_w$ to the high and low temperature structures
as seen in Fig.\,\ref{fig:pdfgui}(c).
In addition to locating the transition itself, the evolution in both temperature 
regimes suggest pure, single-phase polymorphs of the VO$_2$ end members.
There is no broad hysteresis in the transition because the measurement
allows thermal equilibrium to be reached at each point.
In addition, the sample is a bulk powder of large crystallites,
which implies the possibility of many concurrent nucleation events.

\begin{figure}
\centering\includegraphics[width=0.9\columnwidth]{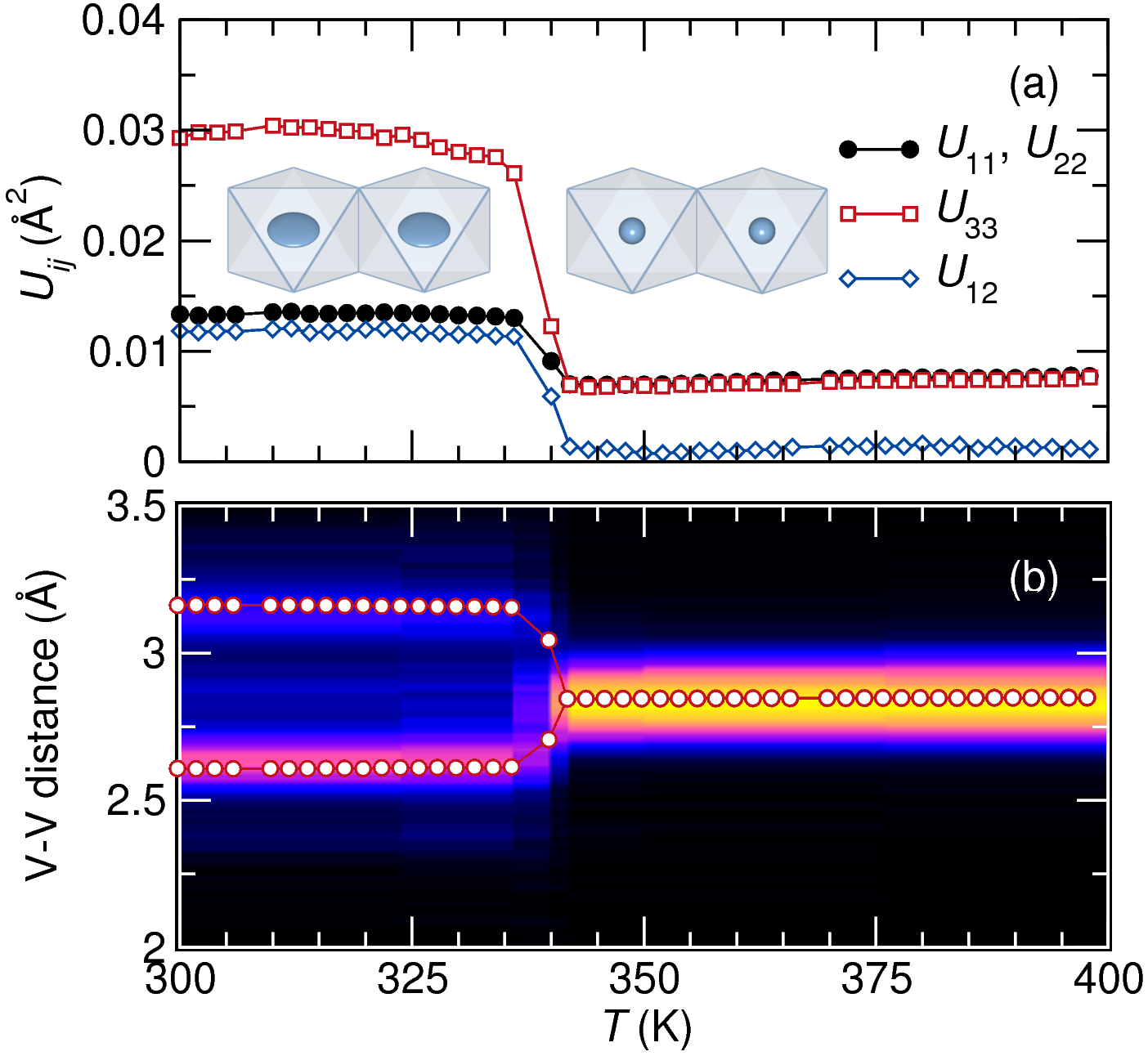} \\
\caption{ (Color online) (a) Thermal parameters ($U_{ij}$) obtained from PDF 
least-squares refinement as a function of temperature reveal abrupt 
changes at the transition temperature. 
Thermal ellipsoids (99\%) for neighboring V polyhedra in the
$c$ direction are shown above and below the transition temperature. 
(b) V--V bond distances from the least-squares refinement 
(points) overlayed with a map of the V--V distances obtained from
RMC modelling.}
\label{fig:uij-vv}
\end{figure}

Dimerization of the V cations leads to splitting of the $2a$ Wyckoff site in the
tetragonal VO$_2$ structure. If a dimerized  structure is fit using the
high-temperature model, the structural changes must be accommodated by an 
increase in the V atomic displacement parameters $U_{ij}$, which 
could convey information of the directionality of the displacement. 
Fig.\,\ref{fig:uij-vv}(a) shows the least-squares refined values of
the $U_{ij}$ parameters of the tetragonal structure over the full
temperature range. At all temperatures, $U_{11} \equiv U_{22}$ by symmetry.
Above the transition, $U_{33}$ is equal to $U_{11}$, while $U_{12}$
(the only non-diagonal parameter allowed by symmetry) is negligible. The
high-temperature V position is therefore spherical, and is shown on the
right in Fig.\,\ref{fig:uij-vv}(a). Below 340 K, dimerization along
the $c$ direction leads to a large increase in $U_{33}$. Tilting of V--V vectors off the
$c$ axis (zero above the transition) produces an increase in
$U_{12}$. The corresponding growth and elongation of the V thermal
ellipsoids are seen on the left in Fig.\,\ref{fig:uij-vv}(a).

The split V positions and corresponding dimerized V--V distances
are the hallmark of cooling through the transition, and we use the PDF
to extract the distances directly.
If there were a gradual shift in V--V distances over the extent of a
wide transition region, which other probes such as the electrical
resistivity suggest,
the local structure information in the PDF would reproduce these distances
regardless of whether or not they are correlated over long ranges. 
Least-squares refinements of the V--V distances (using the appropriate
monoclinic or tetragonal model for each structure) are displayed as points in 
Fig.\,\ref{fig:uij-vv}(b) and show an abrupt change upon cooling through the 
transition. The 
distances at 336\,K and 342\,K are effectively unchanged from the low-
and high-temperature structures at 300\,K and 400\,K respectively, so the 
structural transition occurs within this range ($\Delta T < 6$\,K).
The least-squares refinements, however, are predicated on the choice of a
monoclinic or tetragonal unit cell.

Histograms of V--V distances from RMC supercells (not constrained by
symmetry) are displayed as an intensity map in Fig.\,\ref{fig:uij-vv}(b). 
They display excellent agreement with the least-squares refined values for 
all points except $T$ = 340\,K, implying that the
single-phase least-squares refinements accurately reproduce the true distribution of
V--V  distances in the end members. However, at $T$ = 340\,K the 
least-squares refined distances are of intermediate length, while the RMC V--V
histogram is unclear. Is there an 
intermediate phase with V--V distances that are distinct from 
the monoclinic and tetragonal structures, or is there
a mixture of two phases? Our subsequent analysis of the linear combination of
experimental PDFs, along with V positions from RMC simulations, can resolve this issue.

\begin{figure}
\centering\includegraphics[width=0.9\columnwidth]{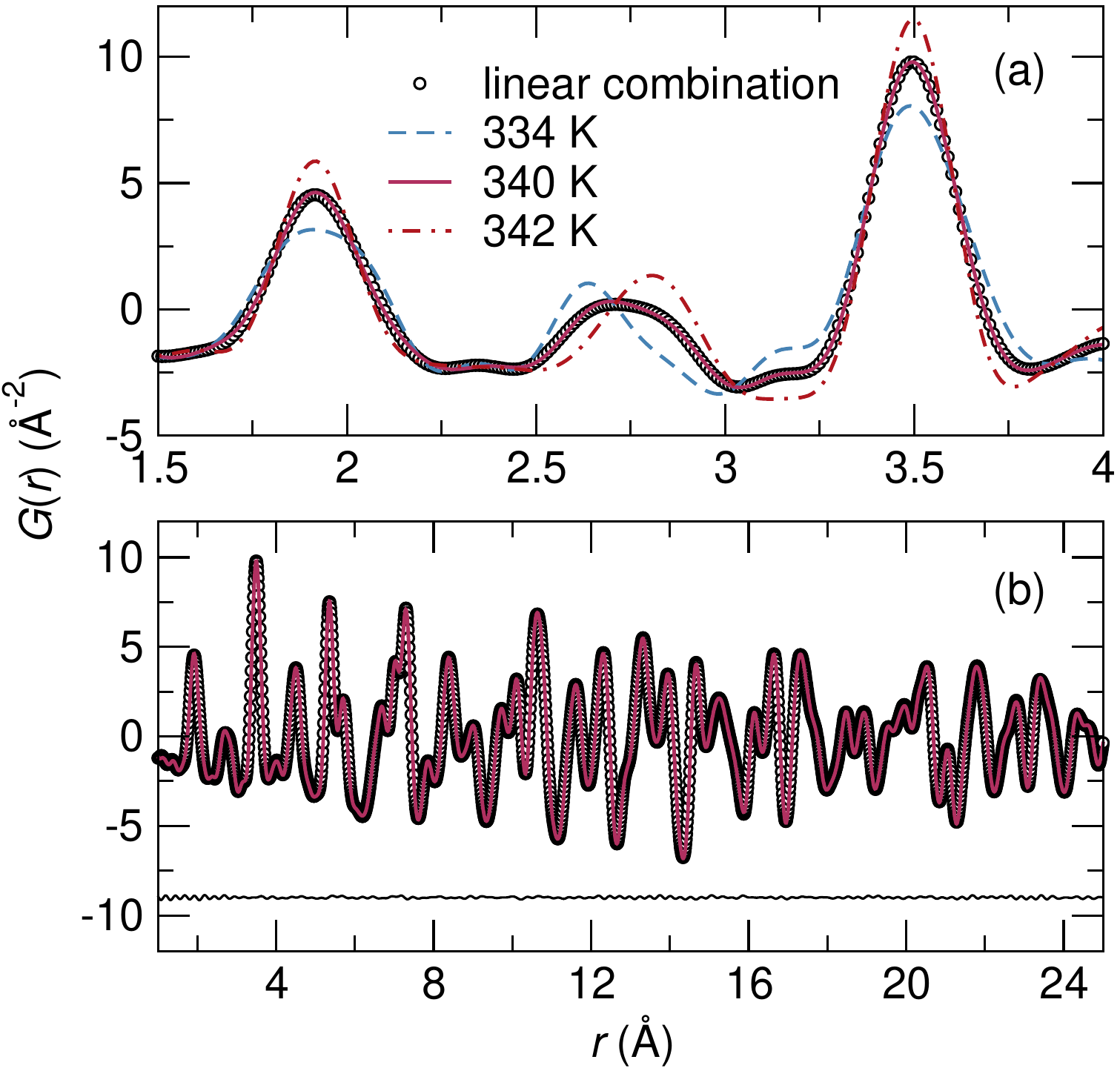} \\
\caption{(Color online) (a) Experimental X-ray PDFs from VO$_2$ at $T$ below
(334\,K), equal to (340\,K), and above (342\,K) the  transition temperature.
A linear combination of the 334\,K and 342\,K data is shown as points,
along with a difference curve to the 340\,K data. In (b) the full fitting
range is shown--at the transition temperature the sample is comprised of
two distinct end-member phases.}
\label{fig:linearcombo}
\end{figure}

In Fig.\,\ref{fig:linearcombo}(a), experimental PDFs are displayed for 
$T$ = 334\,K, 340\,K, and 342\,K: below, during, and above the
structural transition temperature. The $T$ = 334\,K and 342\,K PDFs can be 
refined to the end member structures, so a linear combination of these two 
PDFs produces a two-phase PDF containing three V--V distances: two monoclinic 
and one tetragonal. This linear combination PDF is displayed as the points
behind the $T$ = 340\,K data, and tracks the data up to 25\,\AA. The
difference  curve between the combination PDF and the $T$ = 340\,K data is
shown in Fig.\,\ref{fig:linearcombo}(b), showing that the two PDFs are
identical within the limits of experimental noise. The agreement 
implies that, even on a local scale, the VO$_2$ sample at 340\,K 
comprises two structurally distinct regions, corresponding
directly to the tetragonal or the monoclinic structures. No 
additional contribution to the PDF is found.

\begin{figure}
\centering\includegraphics[width=0.9\columnwidth]{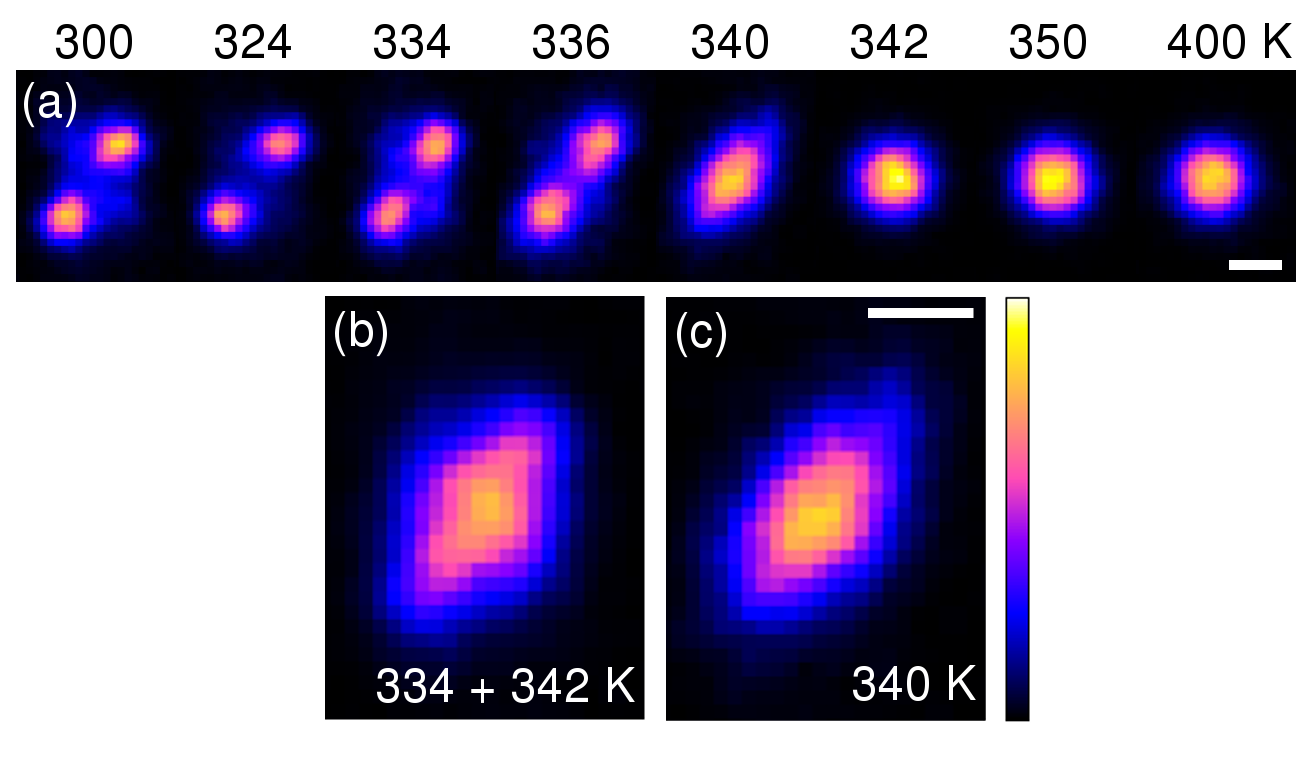} \\
\caption{(Color online) (a) The relative probabilities of V atom positions as obtained from
RMC modeling are dimerized at four temperatures below
and spherical at three temperatures above the transition. At $T$ = 340\,K, the probability 
distribution is intermediate. In (b), a two-phase model is formed by 
averaging the experimental maps from $T $= 334\,K and 342\,K, both of which
resemble the low- and high-temperature end members. This resulting map appears
similar similar to the 340\,K data reproduced in (c). 
Scale bars are 0.2\,\AA. The color bar indicates minimum and maximum 
probability.}
\label{fig:vclouds}
\end{figure}

Corroborating evidence for the two-phase combination is seen in RMC
simulations. We examine the distribution of V positions  around the ideal
crystallographic site in real space by folding the positions of all V
in the RMC supercell back into a single unit cell, which produces a
cloud of 1600 V positions (from a $10 \times 10 \times 16$ supercell) on
each site. This cloud is viewed as a two-dimensional histogram showing the most probable 
positions of V cations in the $a-c$ plane in Fig.\,\ref{fig:vclouds}. 
For each cloud with $T >$ 340\,K, V distributions are spherical (as
seen in the ellipsoids in Fig.\,\ref{fig:uij-vv}(a)) and centered
on the ideal tetragonal positions. For $T <$ 340\,K, dramatic splitting
of the V position is seen. These two spots correspond to the end member
monoclinic positions. We see no
evolution of V clouds versus temperature far from the transition, but the
340\,K cloud possesses an intermediate shape, distinct from the two
regimes on either side.  Just as was performed for the experimental PDFs 
themselves, we can linearly combine RMC results to produce models of a 1:1
mixture of the two phases.  In Fig.\,\ref{fig:vclouds}(b), an average of the 
clouds from 334\,K and 342\,K is displayed, as a model of coexistence of
the end members. The experimental cloud from $T$ = 340\,K is reproduced in
Fig.\,\ref{fig:vclouds}(c), and their similar appearance indicates that this is a
two-phase mixture, in agreement with the linear combinations of the PDF in
Fig.\,\ref{fig:linearcombo}.


Qazilbash \textit{et al.}\cite{Qazilbash_Science} have recently suggested the 
formation of  ``metallic nanopuddles'' at the transition temperature, and
Cava and coworkers\cite{Cava} have suggested the importance
of metal-metal dimers in a series of substituted V$_{1-x}M_x$O$_2$ 
($M$ = Nb, Mo) samples.  What has not been clear up to now is  whether these 
implied a third, intermediate phase of VO$_2$.  Here, we have shown that at 
the transition temperature there are only two distinct phase populations: low 
temperature monoclinic VO$_2$ together with the high temperature tetragonal 
phase. Above the transition, there is no evidence for the presence of V--V 
dimers. 


It is a pleasure to acknowledge R. J. Cava for suggestions and encouragement,
and P. J. Chupas and K. W. Chapman for assistance with data 
collection at APS beamline 11-ID-B, supported by the DOE Office of Basic 
Energy Sciences under contract W-31-109-Eng.-38. SAC acknowledges research 
support from the University of Kent. We received support from the LANL-UCSB Institute 
for Multiscale Materials Studies and the NSF through 
a Career  Award (DMR 0449354) to RS and MRSEC facilities (DMR 0520415). 
RMC simulations were performed using computational resources of the California 
NanoSystems Institute, supported in part by Hewlett-Packard. 

\bibliography{vo2_pdf}

\end{document}